# All-Electrical Control of a Hybrid Electron Spin/Valley Quantum Bit in SOI CMOS Technology

Léo Bourdet, Louis Hutin, *Member, IEEE*, Benoit Bertrand, Andrea Corna, Heorhii Bohuslavskyi, Anthony Amisse, Alessandro Crippa, Romain Maurand, Sylvain Barraud, Matias Urdampilleta, Christopher Bäuerle, Tristan Meunier, Marc Sanquer, Xavier Jehl, Silvano De Franceschi, Yann-Michel Niquet, Maud Vinet

*Abstract*— We fabricated Quantum Dot (QD) devices using a standard SOI CMOS process flow, and demonstrated that the spin of confined electrons could be controlled via a local electrical-field excitation, owing to inter-valley spin-orbit coupling. We discuss that modulating the confinement geometry via an additional electrode may enable switching a quantum bit (qubit) between an electrically-addressable valley configuration and a protected spin configuration. This proposed scheme bears relevance to improve the trade-off between fast operations and slow decoherence for quantum computing on a Si qubit platform. Finally, we evoke the impact of process-induced variability on the operating bias range.

## I. INTRODUCTION

BY leveraging the phenomena of quantum superposition and entanglement, some specifically designed quantum algorithms [1] can achieve polynomial to exponential speed up when compared to their best classical counterparts, thus holding great promise for a variety of applications such as secure data exchange, database search, machine learning, and simulation of quantum processes. Quantum computers are envisioned as hybrid devices [2] where quantum cores operate in conjunction with classical circuitry, part of which is dedicated to programming, control and post-processing functions. While the engineering challenges span across various fields such as physics, electronics, computer science and computer engineering [3], we focus here on the matter of integrating qubits with long coherence times and high-fidelity operations.

The first of DiVincenzo's criteria [3] for a physical implementation of a quantum computer is the ability to define two-level quantum-mechanical systems, and several candidates have emerged in the past decades. Solid-state qubits which can be controlled electrically are generally thought to be more scalable and their manipulation can be performed at the GHz timescale, though it comes at the cost of shorter coherence times. Among the latter, superconducting qubits have been historically leading the race in the implementation of quantum logic. These are however macroscopic objects and as such prone to coupling to probes and environment. Spin qubits, in which the quantum information is encoded in the spin degree of freedom of one [5] or several [6],[7] charged particles, may offer a good compromise owing to their microscopic dimensions. Silicon spin qubits in particular have recently emerged as a promising option, first due to the recent observation of long coherence times and high fidelity [8]-[10], and second thanks to their compatibility with state-of-the-art technologies perfected over several decades by the IC manufacturing industry.

Regarding the first point, the latest notable achievements are the demonstration of single qubit Gates with >99% fidelity [10],[11], and the implementation of quantum algorithms on a two-qubit processor [12]. Fig. 1 shows the evolution of a figure of merit sometimes called "Q-factor" for experimental realizations in the relatively recent area of Si spin qubits.

Maximizing the Q-factor is critical to performing robust calculations since it sets an upper bound to the number of operations that can be sequentially performed on a number of qubits for the implementation of quantum error detection protocols. Whereas increasing the coherence time in a quantum system is generally carried out by further insulating the qubit from its environment, fast manipulation depends on a strong coupling to the excitation signal. The following proposal [18],[19] aims at alleviating this trade-off by making an otherwise protected qubit sensitive to locally applied electrical stimuli solely during the manipulation phase. This manuscript is an extended version of [19], with additional considerations regarding the impact of surface roughness and film thickness variability on finding the proper bias conditions for each device.

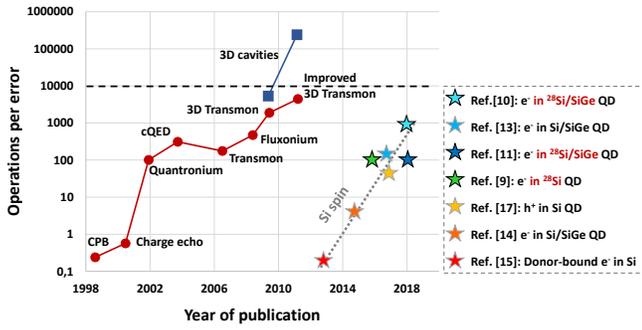

Fig. 1. Graph adapted from [16], showing the number of operations per error for various implementations of superconducting qubits vs. publication year. The added star symbols represent recent demonstrations of Si spin qubits in electrostatically-defined Quantum Dots. The figure of merit was estimated as the dephasing time $T_2^*$ divided by the time needed to induce a π-rotation of the spin. The dashed line corresponds to an error threshold compatible with the implementation of most Quantum Error Correction Codes.

## II. DEVICE AND DEFINITION OF QUANTUM STATES

In recent prior work, we have demonstrated two-axis control of the first hole spin qubit in Si transistor-like structures using a CMOS technology platform [17]-[21]. The first step is the ability to isolate and confine a charged particle, electron or hole, in a Quantum Dot (QD). Our approach consists in using accumulation field-effect Gates to define the confinement potential under e.g. a Si/SiO$_2$ interface. Lateral definition is assisted by mesa patterning of the Si active area. Carrier reservoirs are formed by ion implantation and coupled to the QDs. In this work, the fabrication only differs from a standard CMOS process flow by the deposition of larger SiN spacers with respect to the case of classical devices (30 nm vs. typically ~10 nm or less). They are designed to protect the SOI film from self-aligned doping between dense Gates (64nm pitch), thus leading to a linear arrangement of wrap-around Gates along an intrinsic NanoWire (Fig. 2) [22].

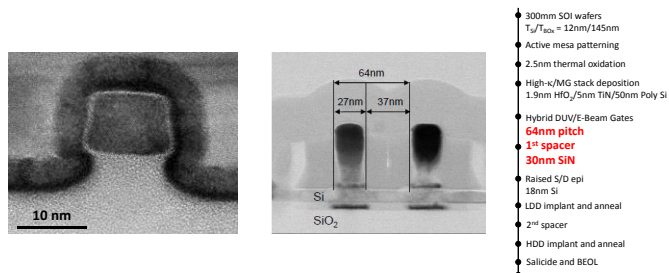

Fig. 2. Left: STEM view along the Gate wrapping around the Si channel. Center: STEM view of two Gates in series (64 nm pitch) showing the width of the 1$^{st}$ spacer. Right: simplified process flow.

At very low temperatures (~1K and below), each Gate defines a QD with a discrete energy spectrum, which can be used to confine a small number of charges controlled by the Coulomb blockade effect (Fig. 3). The wide spacers over an undoped thin film provide tunnel junctions separating the QDs from the charge reservoirs and from one another. Making a qubit out of a QD entails the ability to initialize and manipulate a two-level quantum state of a single charge, such as spin-down $|\downarrow\rangle$ and spin-up $|\uparrow\rangle$. Spin degeneracy can be lifted by means of an externally-applied static magnetic field B, the so-called Zeeman splitting energy being $E_Z=|g|.\mu_B.B$ where g is the Landé g-factor (g≈2 for electrons in Si) and $\mu_B$ the Bohr magneton.

Initialization is simply performed by waiting for the system to relax to its ground state $|\downarrow\rangle$.

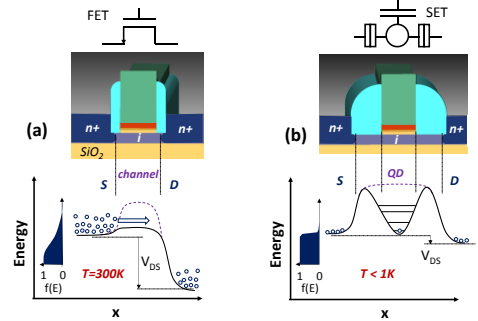

Fig. 3. Energy profile along the channel of (a) a SOI FET at 300K in which carriers flow continuously above a lowered barrier (b) a SET operating in the Coulomb Blockade regime at low T due to large tunnel barriers beneath the spacers and energy quantization in the Gate-defined Quantum Dot (QD).

Inducing transitions by Electron Spin Resonance (ESR) with an RF magnetic-field is the most straightforward approach to spin control (Fig. 4), but the excitation can hardly be applied locally. This can be a drawback for maximizing the manipulation speed, which depends on the coupling strength. Electrical Dipole Spin Resonance (EDSR) driven by a Field-Effect Gate placed directly on top of the qubit can in principle be achieved with the assistance of a micromagnet producing a magnetic-field gradient in the vicinity. This causes the particle traveling back and forth to perceive an oscillating B-field [10]-[14],[23]. However, this approach can be demanding in terms of integration and design of large-scale qubit arrays. A more compact solution would be to rely on the intrinsic Spin-Orbit Coupling (SOC) of the material(s), which tie the spin of a particle to its orbital motion and hence all-electrical oscillating signals. Unfortunately in Si electrons, unlike holes, have generally weak intrinsic SOC.

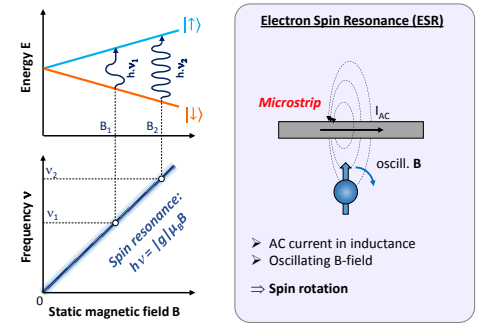

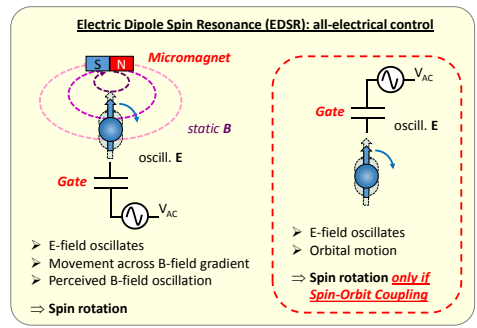

Fig. 4. (Top left) Principle of Zeeman splitting between $|\downarrow\rangle$ and $|\uparrow\rangle$, resonant transitions and spin resonance signature. (Top right) B-field-mediated ESR. (Bottom) E-field-mediated EDSR, either relying on intrinsic Spin-Orbit Coupling (SOC), or using a micro-magnet as an auxiliary.

Our test device for EDSR demonstration (Fig. 5) consists of a Two-Gate nFET-like structure with Gates partially wrapping around the [110]-oriented SOI NanoWire (W=30nm; H=12nm) [25]. As shown in [27], electron localization occurs along the upper edges of the mesa. We consider two QDs, QD1 and QD2 confined in the "corners" defined by gates $G_1$ and $G_2$. If both are in the same spin state (*e.g.* parallel spins, which is the ground state in a finite magnetic field B), Pauli's exclusion principle prevents charge movement from QD1 to QD2, and hence $I_{DS}$ current from flowing. However, a spin rotation obtained by applying a resonant RF E-Field to $G_1$ would lift the Pauli Spin Blockade and enable a non-zero current.

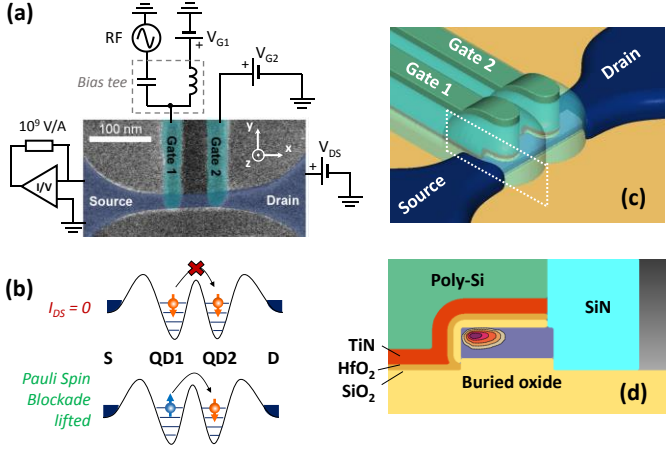

Fig. 5. (a) Top-view SEM of the two-Gate device after Gate patterning, and setup description. (b) Spin-filtering mechanism across the Double QD based on the Pauli Spin Blockade rectifying the Drain current. (c) Schematic view of the partially wrapping Gates. (d) Cross-section along a Gate and representation of the asymmetrical electron wavefunction along the mesa edge, or "Corner Dot".

Yet for electrons in Si, the additional valley degree of freedom needs to be considered. The conduction band of bulk Si features six degenerate Δ valleys. Structural and electrical confinement in our device, however, leaves two low-lying valleys $v_1$ and $v_2$, separated by an energy $\Delta_V$. From these two valleys, four distinct states can be resolved upon applying a static magnetic field: $|v_1, \downarrow\rangle$, $|v_1, \uparrow\rangle$, $|v_2, \downarrow\rangle$ and $|v_2, \uparrow\rangle$.

### III. CORNER DOTS AND SPIN-VALLEY MIXING

Of particular interest are the two states $|v_1, \uparrow\rangle$ and $|v_2, \downarrow\rangle$, which may be mixed under the condition that an inter-valley spin-orbit (SO) coupling coefficient $C_{v1v2}$ in the Hamiltonian is non-zero. As illustrated in **Fig. 7**, this criterion is fulfilled if the mirror symmetry of the electron wavefunction with respect to the (XZ) plane is broken [24],[25]. The partially overlapping Gate leading to the "Corner Dot" confinement is therefore the key to spin-valley-orbit mixing in this case.

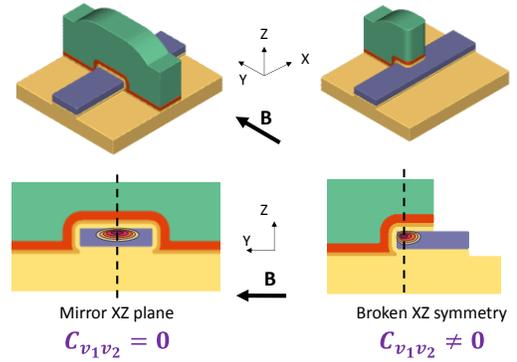

Fig. 6. Impact of device geometry on inter-valley Spin-Orbit Coupling. The coupling term $C_{v1v2}$ is non-zero if the symmetry of the electron wavefunction with respect to the (XZ) plane is broken. This condition is fulfilled in the case of Corner Dots.

As B is increased and the spin splitting $E_Z=|g|.\mu_B.B$ approaches the valley splitting $\Delta_V$, the $|v_1, \uparrow\rangle$ and $|v_2, \downarrow\rangle$ energies may either cross (no coupling) or anticross ($C_{v1v2} \neq 0$). In the former case (Fig. 7a)), only spin-preserving inter-valley transitions can be expected in response to pure E-field excitations. In the latter case, due to states mixing near the anticrossing, B-dependent spin/valley transition diagonals may add-up to the EDSR signal (Fig. 7b)). A color plot of $I_{DS}$ measured in a dilution cryostat at T=15mK vs. E-field frequency and B clearly shows spin resonance lines (Fig. 8). This is to our knowledge the first experimental measurement of micromagnet-free resonant E-field manipulation of single electron[1] spins in Si QDs [25].

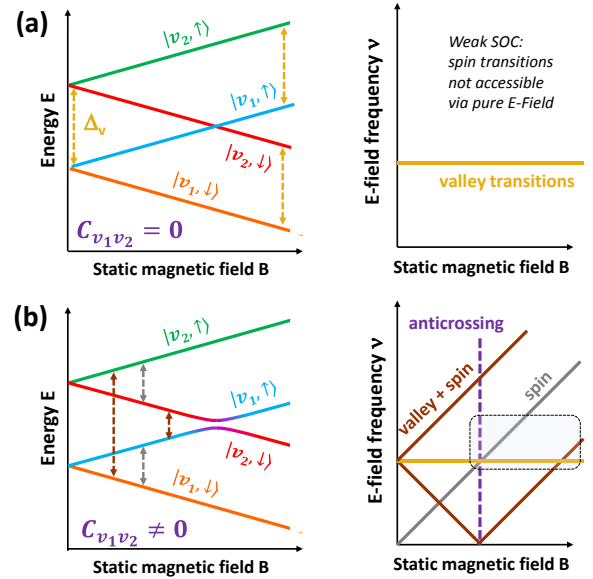

Fig. 7. (a) Zeeman splitting from $v_1$ and $v_2$ in the case of no inter-valley SOC, and associated expected EDSR signal. (b) Case in which inter-valley SOC exists and states anti-cross, and expected EDSR. The dotted frame symbolizes the region measured in Fig. 8.

---

[1] Magnet-free resonant E-field manipulation of a two-electron qubit in the $\{|\uparrow\downarrow\rangle, |\downarrow\uparrow\rangle\}$ basis was recently demonstrated in [26], by leveraging the g-factor difference between two quantum dots ($f_{AC}$ = 0.9MHz ≈ $|\Delta g|.\mu_B.B/h$ ).

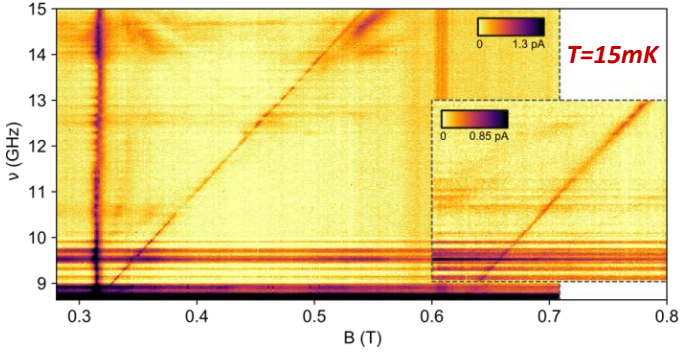

Fig. 8. Experimentally measured EDSR signal (two measurements combined, with insert for higher magnetic fields) in the dotted frame region of Fig. 7 (b), showing spin and spin/valley transitions.

## IV. PROGRAMMING A VALLEY STATE, ENCODING A SPIN STATE

Since the splitting between $v_1$ and $v_2$ is related to charge confinement close to an interface, it is possible to tune $\Delta_V$ by modulating the vertical electric field. This was shown in [28] using coplanar side Gates on bulk Si, but SOI offers the possibility of using the Back-Gate potential $V_b$. We calculated the $\Delta_V(V_b)$ energy dependence using a Tight Binding model for the valley and the SO coupling at the atomistic level [29]. The results are shown in Fig. 9 together with corresponding plots of the electron wavefunction. The tunability of $\Delta_V$ can be leveraged as schematized on Fig. 10: adiabatically changing $V_b$ allows following the lower branch past the anticrossing and transitioning continuously from $|v_1,\uparrow\rangle$ to $|v_2,\downarrow\rangle$. If one defines the qubit basis states $|0\rangle$ as $|v_1,\downarrow\rangle$ and $|1\rangle$ as this hybridized lower branch, $V_b$ enables to switch between a pure spin regime and a pure valley regime. The advantage of a valley qubit is the all-electrical addressability of inter-valley transitions, the downside being sensitivity to charge noise and hence shorter decoherence times. Conversely, when in spin regime, the qubit is scarcely addressable electrically but benefits from a longer lifetime.

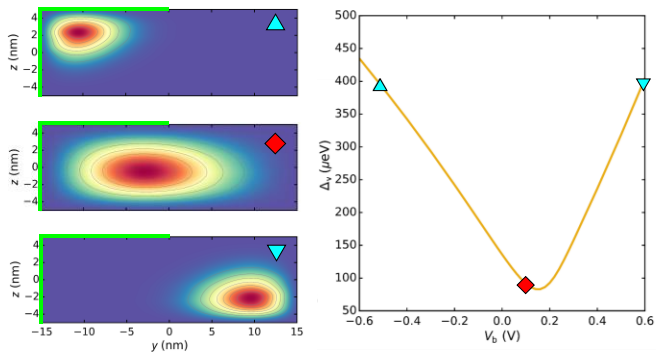

Fig. 9. Simulated influence of the SOI Back-Gate voltage $V_b$ on the $\Delta_V$ valley splitting for an ideal device (no surface roughness). $\Delta_V$ is maximal when the charge is confined against an interface. At first, a more positive $V_b$ tends to pull the wavefunction towards the center of the NanoWire, away from the interfaces. A further $V_b$ increase results in increasing $\Delta_V$ again, due to charge confinement against the interface with the buried oxide. Here the width and height of the wire are W=30 nm and H=10 nm, the gate is 30 nm long, and the buried oxide is 25 nm thick. The y and z position of the Front Gate, biased at $V_{fg}=+0.1V$, is highlighted in green.

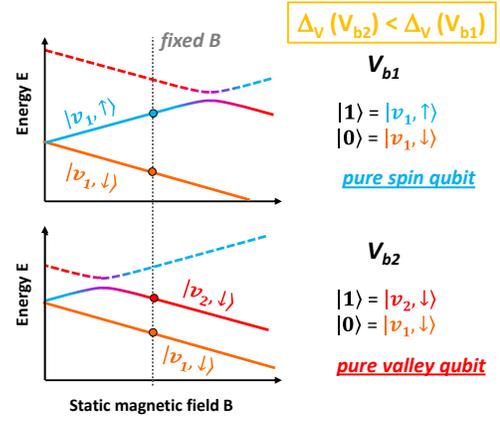

Fig. 10. Energy diagram showing two $V_b$ configurations. At a given B, changing $V_b$ adiabatically enables to switch between a spin qubit and a valley qubit regime, by operating respectively left and right of the anti-crossing.

This approach leads to circumventing a trade-off between qubit manipulation speed and coherence time, thus improving the number of operations/error. Advantageously, the qubit rotation speed is maximal when the charge is pulled away from the interfaces, which is more challenging to achieve by using only coplanar Front Gates as proposed in [30]. Fig. 11 shows the simulated chronograms of the electrical RF Gate 1 excitation signal ($\nu$ = 23.66 GHz), the resulting Rabi oscillations of the qubit ($f_{Rabi}$ = 80 MHz) in valley mode, and the eventual spin rotation as $V_b$ adiabatically ramps past the anticrossing back to spin mode.

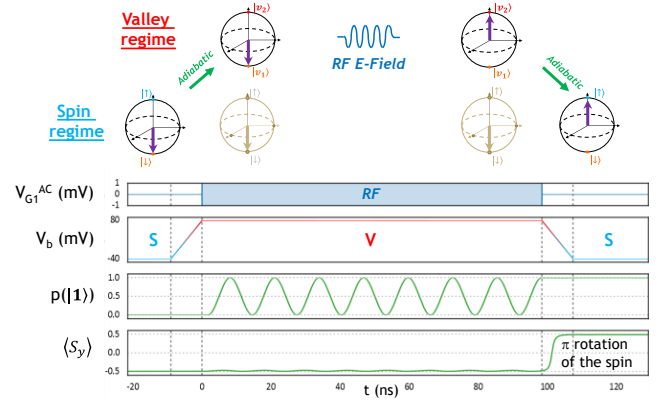

Fig. 11. Simulated purely electrical manipulation of the spin of a confined electron. A $V_b$ ramp brings the qubit in the valley regime, in which it can oscillate ($f_{Rabi}$ = 80MHz) in response to an RF E-field excitation (here $\nu$=23.66 GHz). As the $V_b$ ramp is reversed, the $|1\rangle$ eigenstate transitions from $|v_2,\downarrow\rangle$ to $|v_1,\uparrow\rangle$, thus leading to a $\pi$ rotation of the spin. p($|1\rangle$) is the probability to be in the $|1\rangle$ state and $\langle S_y \rangle$ is the average spin along $y$ (the direction of the magnetic field).

## V. DISCUSSION

In this section, we provide some insight on the impact of process variations on the optimal operating range of the back-Gate bias-mediated scheme described above. We have re-calculated $\Delta_V(V_b)$, accounting for local variability due to surface roughness (Fig. 12). The surface roughness profiles are generated from a Gaussian auto-correlation function with rms=0.4 nm consistent with room-temperature mobility measurements in similar devices. It can be noted that the $V_b$ value where the Zeeman and valley splitting match, *i.e.* the anticrossing point, can be spread over as much as 0.5V. The most negative $V_b$ value should serve as an upper bound for the "hold" operation in spin regime. For "program", $V_b$ should be targeted near the minimum of $\Delta_V$ where manipulation is the fastest and takes place at the resonance frequency $\Delta_V/h$. This corresponds to a regime in which the electron is pulled away from both interfaces and is thus least affected by surface roughness, hence the smaller dispersion.

The strength of inter-valley Spin-Orbit coupling $C_{v1v2}$, which ultimately determines the manipulation speed, displays a similar trend versus $V_b$ as the valley splitting (Fig. 14). This is because a reduction of vertical confinement coincides with a recovery of the wavefunction symmetry. It is however noteworthy that $C_{v1v2}$ is much more robust against interface disorder, a trend which is confirmed when varying film thickness H.

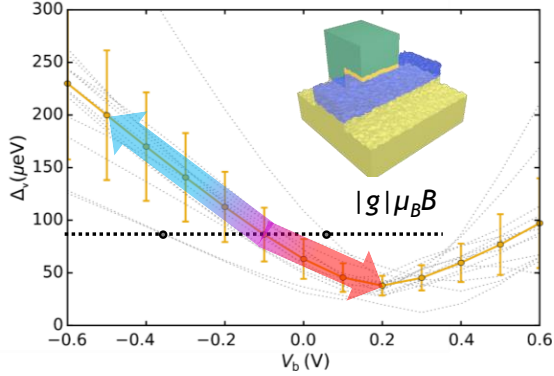

Fig. 12. Impact of local surface roughness variability on the $\Delta_V(V_b)$ dependence (rms 0.4 nm). The spreading tends to be less severe near the $\Delta_V$ minimum, so the magnetic field can be chosen to operate close to this point. As $\Delta_V=|g|.\mu_B.B$ defines the anticrossing point, traveling up the curve leads to the spin regime, and down to the valley regime.

Figure 13 shows the dependence of the minimum valley splitting $\Delta_V^{min}$ near $V_b=0.2$ V as a function of the height H of the nanowire. Different realizations of surface roughness disorder (dotted lines) are averaged (solid line). $\Delta_V^{min}$ decreases as the height of the nanowire increases and the wave function can be further decoupled from the top and bottom surfaces. The remaining oscillations are the fingerprints of the interferences between wave reflections on the top and bottom interfaces which are not completely washed out by surface roughness. The minimum valley splitting can be brought well below 75 µeV in devices with H > 8 nm, so that the qubit can be manipulated at reasonable frequencies $\Delta_V/h$<20 GHz.

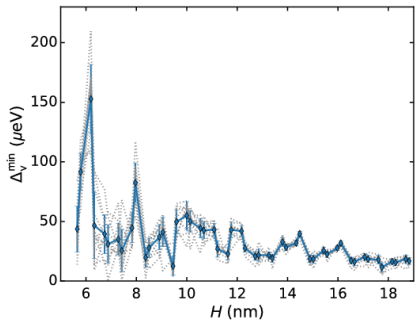

Fig. 13. Dependence of the minimum valley splitting $\Delta_V^{min}$ near $V_b=0.2$ V as a function of the height H of the nanowire. Each dotted line is a separate surface roughness sample; the solid line is the average.

In practice, it may be desirable to control all qubits at a unique resonance frequency $\nu<|g|.\mu_B.B/h$ and adjust $V_b$ so that $\Delta_V=h\nu$ matches that frequency. This however calls for individual back gates and for a calibration of each qubit device.

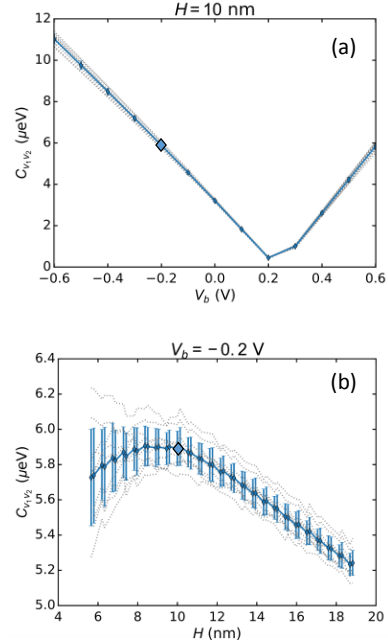

Fig. 14. (a) Impact of local surface roughness variability for a device with H=10 nm on the intervalley SOC coefficient $C_{v1v2}$ as a function of $V_b$. $C_{v1v2}$ is extremely robust with respect to surface roughness, so that SOC is always significant at the anticrossing point. $C_{v1v2}$ reaches a minimum around $V_b=0.2$ V, corresponding to an additional approximate symmetry plane. (b) $C_{v1v2}$ as a function of H, near the anticrossing region at $V_b=-0.2$ V. $C_{v1v2}$ is again extremely robust, even on large variations of the nanowire thickness.

## VI. CONCLUSIONS

We observed spin transitions in MOS Gate-confined electrons in a Si NW using only E-field excitations and without resorting to co-integrated micromagnets. The underlying mechanism is based on the interplay between Spin-Orbit Coupling (SOC) and the multi-valley structure of the Si conduction band, and is enhanced by the "Corner Dot" device geometry. By offering the ability to break and restore the confinement symmetry at will, the SOI Back-Gate permits fast programming in valley mode, and information storage in spin mode. This functionality could alleviate the trade-off between fast manipulation and long coherence time, thereby improving the outlook for compact, scalable and fault-tolerant quantum logic circuits. Considering the valley-splitting-dependent resonance frequency for driving coherent oscillations of the qubit, it is probable that separate back-Gates should be defined in order to calibrate each device to a common operating point.

ACKNOWLEDGMENT

The authors acknowledge financial support from the EU under Project MOS-QUITO (No. 688539) and the Marie Curie Fellowship within the Horizon 2020 program, and from the French National Research Agency (ANR) through projects IDEX UGA (ANR-15-IDEX-0002) and CMOSQSPIN (ANR-17-CE24-0009). Part of the calculations was run on the TGCC/Curie and CINECA/Marconi machines using allocations from GENCI and PRACE.